\documentclass[12pt,a4paper]{article}
\usepackage{amsmath,amssymb, amsfonts, amsthm}
\usepackage{authblk}

\begin{document}

\title{On the equivalence of Clauser-Horne and 
Eberhard inequality based tests}

\author[1]{Andrei Khrennikov\footnote{email: Andrei.Khrennikov@lnu.se}}
\author[2,3,4]{Sven Ramelow}
\author[2]{Rupert Ursin}
\author[2,3]{Bernhard Wittmann}
\author[5]{Johannes Kofler}
\author[6]{Irina Basieva}
\affil[1]{International Center for Mathematical Modeling \\
in Physics, Engineering, Economics, and Cognitive Science\\
Linnaeus University, V\"axj\"o-Kalmar, Sweden}
\affil[2]{Institute for Quantum Optics and Quantum Information – Vienna (IQOQI), Austrian Academy
of Sciences, Boltzmanngasse 3, Vienna, Austria}
\affil[3]{Quantum Optics, Quantum Nanophysics, Quantum Information, University of Vienna, Faculty
of Physics, Boltzmanngasse 5, Vienna, Austria}
\affil[4]{Cornell University, 271 Clark Hall, 142 Science Dr., Ithaca, 14853 NY, USA}
\affil[5]{Max Planck Institute of Quantum Optics (MPQ), Hans-Kopfermann-Str. 1, 85748 Garching, Germany}
\affil[6]{Prokhorov General Physics Institute,  Vavilov str. 38D, Moscow, Russia}

\maketitle

\begin{abstract} Recently, the results of the first experimental test for entangled photons closing the 
detection loophole (also referred to as the fair sampling loophole)
were published (Vienna, 2013).  From the theoretical viewpoint the main  
distinguishing feature of this long-aspired 
experiment was that the 
Eberhard inequality  was used. Almost simultaneously another experiment closing this loophole was performed 
(Urbana-Champaign, 2013) and it was based on the Clauser-Horne  inequality (for probabilities). 
The aim of this note is to analyze the mathematical and experimental equivalence of tests 
based on the Eberhard inequality and various forms on the Clauser-Horne inequality.  
The structure of the mathematical equivalence is nontrivial. In particular, it is necessary to 
distinguish between algebraic and statistical equivalence.  Although the tests based on these 
inequalities are algebraically equivalent,   
they need not be equivalent statistically, i.e.,
theoretically the level of statistical significance can drop 
under transition from one test to another (at least for finite samples). Nevertheless, 
the data collected in the Vienna-test 
implies not only a statistically significant violation 
of the Eberhard inequality, but also of the  Clauser-Horne inequality (in the ratio-rate form):
 for both a violation $>60\sigma.$
\end{abstract}

\section{Introduction}

Experimental realization of a loophole-free test for Bell \cite{B} inequalities will have impact both for quantum foundations 
and quantum technologies. In both cases the present situation,  
e.g., in quantum cryptography \cite{Pearle} and 
quantum random generators \cite{RG1} (see also \cite{RG2} for discussion)  is unsatisfying from the scientific viewpoint. 
To experimentally falsify local realism, 
a so called loophole-free Bell experiment will have to be
accomplished successfully.\footnote{We remark that a priori one still cannot exclude 
the possibility that in the final loophole free experiment 
the Bell inequality would be satisfied. In such (very improbable) case, since quantum theory predicts that, for the state under 
preparation, Bell's inequality has to be violated, the experiment would imply rejection of 
the quantum model.  Thus the Bell test can also be considered as an attempt to falsify quantum 
mechanics. (At the initial stage of Bell experimentation expectation that quantum mechanics 
 would be falsified was quite common.)} 
 This was not claimed so far in any of the reported experiments.
Up to now, space-like separation of measurements and basis choices has been accomplished in the pioneering experiments of Aspect {\it et al.} 
 \cite{Aspect, ASP1} and  Weihs {\it et al.}  \cite{Weihs}, closing the so-called locality loophole.\footnote{Although 
we discuss only Bell  tests for entangled photons, it is relevant that the first closure of the detection loophole was achieved with massive particles \cite{ROWE}.} 
These  experimental  tests were based on the 
Clauser-Horne-Shimony-Holt (CHSH) inequality \cite{CHSH}. 
For this inequality
to falsify local realism one must either approach very high total detection efficiency (which includes the optical losses of the setup 
and the efficiency of the detectors) $\eta = 82.8\%,$
or proceed under an assumption to circumvent the loss -- the so-called fair sampling assumption, see section \ref{Fair} for discussion.
It has been shown that using unfair sampling at sufficiently low detection efficiencies, 
very simple local models with hidden variables can violate the CHSH-inequality, e.g., \cite{Pearle}, \cite{CH74}-- \cite{KH_JA}.
Thus in the CHSH-framework, approaching very high detection efficiency is the only possible way to experimentally falsify the classical world view of local realism.        
In spite of technological progress and the existence of detectors whose efficiency nears unity, 
(e.g., TES detectors with efficiency around 95\% \cite{Lita}), it is very challenging to approach the required total detection efficiency.

In 1974, Clauser and Horne \cite{CH74} proposed a new inequality that is not based on 
the fair sampling assumption.  This inequality is expressed in terms of probabilities 
and we shall call it {\it the CH-inequality for probabilities}:
\begin{equation}
\label{CH74_E30}
p(\alpha_1, \beta_1)  + p(\alpha_1, \beta_2)  
+p(\alpha_2, \beta_1) - p(\alpha_2, \beta_2) \leq p^A(\alpha_1) +p^B(\beta_1),
\end{equation} 
where $p(\alpha, \beta)$ and $p^A(\alpha), p^B(\beta)$ are probabilities for coincidence and single counts\footnote{
``Single counts'' are defined as all counts registered on one side for a given setting.}, respectively; see 
section 4 for more detail.

In this paper we shall discuss various forms of the CH-inequality, see section 4 
(see also the review of Clauser and Shimony \cite{CSH}  and the Stanford encyclopedia paper of Shimony \cite{SHI} for details).
Therefore we address each form with the corresponding label. However, we restrict our considerations to the class of 
the CH-inequalities not based on experimentally untestable auxiliary assumptions. Thus we shall not consider the CH-inequality
whose derivation is based on the {\it ``no-enhancement assumption''}\cite{CH74}: if an analyzer is removed from 
one of the paths, the resulting probability of detection is at least as great as with an analyzer.

To determine probabilities in (\ref{CH74_E30}), one has to know the total number of emitted
pairs of photons. As was pointed out by Clauser and Horne \cite{CH74}, see also the review of Clauser
 and Shimony \cite{CSH} for extended discussion,
it is practically impossible to determine this number experimentally. To escape this
problem,  it was proposed \cite{CSH} to exclude the total number of emitted
pairs from consideration by considering a version of the CH-inequality in the form of 
{\it ratio of detection counts rates}\footnote{We remark that 
an important implicit assumption of applicability of 
this inequality is  the assumption of (statistically) constant production rate for pairs of photons, 
see also section \ref{A} for discussion.}: 
\begin{equation}
\label{CH74_R0}
T = \frac{R(\alpha_1, \beta_1)  + R(\alpha_1, \beta_2)  
+R(\alpha_2, \beta_1) - R(\alpha_2, \beta_2)}{R^A(\alpha_1) +R^B(\beta_1)} \leq  1,
\end{equation} 
where $R(\alpha, \beta)$ and $R^A(\alpha), R^B(\beta)$ are coincidence and single rates, respectively;
see section 4 for details.

For the CH-inequality for probabilities (\ref{CH74_E30}), Clauser and Horne \cite{CH74} formulated 
the restrictions on the experimental setup in a straightforward way: first finding the ``optimal angles'' $(\alpha, \beta)$  and then calculating 
other experimental parameters, namely the degree of entanglement and detection efficiency,  to violate the 
inequality, see equations (5) and (6) in \cite{CH74}. It can be shown
that  this procedure leads to very high detection efficiency. To violate the CH-inequality for probabilities,
the detection efficiency has to be at least 82.8\%, see \cite{Pearle, Mermin}.

In \cite{Eberhard} Eberhard proposed a different approach by {\it jointly optimizing} all aforementioned parameters. 
He derived a new Bell inequality  which we will abbreviate by ``E-inequality'':
$$
J\equiv n_{oe}(\alpha_1, \beta_2) +
n_{ou}(\alpha_1, \beta_2)
 + n_{eo}(\alpha_2, \beta_1)+n_{uo}(\alpha_2, \beta_1)+n_{oo}(\alpha_2, \beta_2)
$$
\begin{equation}
\label{FUND}
 - n_{oo}(\alpha_1, \beta_1) \geq 0,
\end{equation}
where $n_{xy}(\alpha_i, \beta_j)$ is the number 
of pairs detected in a given time period for settings $\alpha_i, \beta_j$ with outcomes 
$x,y=o,e, u$ and the outcomes $(o)$ and $(e)$ correspond to detections 
in the ordinary and extraordinary beams, respectively, and
the event that photon is undetected is denoted by the symbol $(u).$  
We point to the main distinguishing features of the E-inequality:

\medskip

a) derivation without the fair sampling  assumption (and without the no-enhancement assumption);

b) taking into account undetected photons;

c) background events are taken into account;

d)  the {\it linear form of presentation} (non-negativity of a linear combination 
of coincidence and single rates).

\medskip

The latter feature (which is typically not emphasized in the literature) 
is crucial to find a simple procedure of optimization of experimental parameters 
and, hence, it makes the E-inequality the most promising experimental 
test to close the detection loophole and to reject local realism without 
the fair sampling assumption. Eberhard's optimization 
has two main outputs which play an important 
role in the experimental design:

\medskip

E1). It is possible to perform an experiment without fair sampling assumption for detection efficiency less than 82,8\%.
Nevertheless, detection efficiency must still be very high, at least 66.6\% (in the absence of background).

E2). The optimal parameters correspond to non-maximally entangled states. 
  
\medskip
   
In 2013, the possibility to proceed with overall efficiencies lower than 82.8\%  (but larger than 66.6\%) 
was explored for the E-inequality and the first  experimental test (``the Vienna test'') closing the detection  
loophole was published \cite{Zeilinger},  for more detailed presentation of statistical data see 
also \cite{Zeilinger2}, \cite{Zeilinger1}.

Almost simultaneously another experiment closing the detection  
loophole was performed \cite{Kwiat1}, \cite{Kwiat2}  based on the probability version of
the CH-inequality \cite{CH74}, see (\ref{CH74_E30}).   

\medskip

In this note we analyze the mathematical and experimental equivalence  of 
the tests based on the E-inequality (\ref{FUND}), and the CH-inequality in 
the {\it ratio-rate form} (\ref{CH74_R0}) -- in fact, its modification  
for the {\it ratio of detection counts}, see section 4, the inequality 
(\ref{CH74_Eo}).  In particular, {\it one has to 
distinguish between algebraic and statistical equivalence} (and for the latter, the cases of  finite and infinite samples).
Although these 
inequalities are (trivially) algebraically equivalent, (see section \ref{ECH}),  
the tests based on them need not be equivalent statistically (for finite samples), i.e.,
theoretically the level of statistical significance can change essentially 
under transition from one test to another, section \ref{stat1}. Nevertheless, 
the data  collected in the Vienna test \cite{Zeilinger}
implies not only the statistically significant violation 
of the E-inequality, but also of the CH-inequality for ratio of detection counts and, hence, for ratio of detection rates: 
for both a violation $> 60 \sigma.$

One of the aims of this note is to determine confidence intervals from the statistics of the data collected in \cite{Zeilinger}.
We remark that if one does not assume that data is Gaussian, then it is impossible to determine the confidence interval exactly 
with the aid of the standard deviation. However, it is possible to estimate it by using the {\it Chebyshev inequality}, see, e.g.,
\cite{Non}. This inequality is applicable under  
the most conservative (worst case) assumption -- namely that the dispersion is finite.
Although the Chebyshev inequality gives only rough estimates of probabilities, in our case (for the data collected in the Vienna
test) it is sufficiently powerful to estimate confidence intervals showing that the hypothesis about the local realistic 
description of the Vienna data must be rejected. 

\section{Fair sampling assumption}
\label{Fair}

The fair sampling assumption plays a crucial role in justification of tests
 based on the CHSH-inequality. In \cite{CHSH} it was stated as  

\medskip

{\it ``if  a pair of photons 
emerges from [the polarizers], the probability of their joint detection is 
independent of [polarizer orientations].''}
 
\medskip

Extended discussions on the role of this assumption in the resolution of the classical-quantum dilemma can be found in the papers 
of Pearle \cite{Pearle} and Clauser and Horne \cite{CH74} and Aspect's PhD thesis \cite{ASP1}, see also Aspect's ``naive experimentalist presentation'' of Bell's 
tests  \cite{ASP2}.  Later the fair sampling assumption was analyzed in  detail in the PhD theses of 
Larsson \cite{L2} and Adenier \cite{AD1}.      

The fair sampling assumption is not made in a Bell test based 
on the E-inequality or any CH-inequality. We also remark that the no-enhancement assumption is not present 
in the list of assumptions for the derivation of the
E-inequality, see \cite{Eberhard}, the assumptions (i)-(iii). 
 
Finally, we remark that in this paper we do not discuss two other important loopholes, {\it the coincidence-time loophole} (and 
the role of space-time in the Bell argument in general,   cf. \cite{Lo}-\cite{Gill}, \cite{KHR_CONT}) and 
the freedom-of-choice loophole \cite{SCheid} (and its relation to the 
impossibility to use the conventional model of classical probability theory, the 
Kolmogorov model, 1933, see \cite{Marian}-\cite{Vor}, \cite{KHR_CONT}, \cite{Hess1, Hess2}).

\section{On equivalence of the E-inequality to the CH-inequalities}

\subsection{E-inequality}
\label{EBB}

We follow Eberhard \cite{Eberhard}: Photons are emitted in pairs $(a,b).$
Under each measurement setting $(\alpha, \beta)$, the events in which the photon $a$ is detected
in the ordinary and extraordinary beams  are denoted by the symbols $(o)$ and $(e)$, respectively, and
the event that it is undetected is denoted by the symbol $(u).$ The same symbols are used to denote the corresponding
events for the photon $b.$ Therefore for the pairs of photons there are nine types of events: $(o, o), (o, u),
(o, e),$ $ (u, o), (u, u), (u, e), (e, o), (e, u),$ and $(e, e).$ 
 (We remark that originally Eberhard considered the experiment with four detectors. In \cite{Zeilinger} his experimental 
design was modified to proceed with only two detectors. We shall come back to this point in section \ref{ECH}.) 

Under the conditions of locality, realism  and statistical reproducibility the inequality (\ref{FUND}), see section 1, was derived.

\subsection{Algebraic equivalence}
\label{ECH}

As was mentioned in section \ref{EBB}, originally Eberhard derived his inequality for the four-detectors experiment, one detector 
at each output of two PBSs. In \cite{Zeilinger} it was shown that the ``$e$-outputs'' can be eliminated from the 
E-inequality (\ref{FUND}), i.e., the four-detectors experimental design can be transformed into the two-detectors design corresponding 
to detection of only  ``$o$-outputs''.  In this section we demonstrate that the latter version of the E-inequality 
is {\it algebraically equivalent} to the CH-inequalities in various forms: for probabilities, ratio of probabilities, 
ratio of rates, and ratio of detection counts. 

As was pointed out \cite{Zeilinger}, the E-inequality can be transformed into the following inequality:
\begin{equation}
\label{CH74_E}
-n_{oo}(\alpha_1, \beta_1) +S_o^A(\alpha_1) - n_{oo}(\alpha_1, \beta_2) +S_o^B(\beta_1) - 
n_{oo}(\alpha_2, \beta_1) + n_{oo}(\alpha_2, \beta_2) \geq 0,
 \end{equation} 
where $S_o^A(\alpha_1)$ and $S_o^B(\beta_1)$ are numbers of single counts in the $o$-channels 
for ``Alice'' and ``Bob'', in settings $\alpha_1$ and $\beta_1$ respectively.   
To match with the CH-inequalities completely, we change the sign and collect singles terms in the right-hand side:
\begin{equation}
\label{CH74_E1}
n_{oo}(\alpha_1, \beta_1)  + n_{oo}(\alpha_1, \beta_2)  
+n_{oo}(\alpha_2, \beta_1) - n_{oo}(\alpha_2, \beta_2) \leq S_o^A(\alpha_1) +S_o^B(\beta_1)  .
\end{equation} 
Then by dividing by the number of emitted pairs $N$ on both sides (and omitting the index ``o'') we obtain the {\it CH-inequality 
for probabilities}.\footnote{This inequality is sometimes referred simply as ``CH-inequality'', see Remark 1 for a short 
discussion on the terminology 
related to the papers of Clauser and Horne \cite{CH74} and Clauser and Shimony \cite{CSH}.} 
(Here we proceed under the assumption of statistically constant production rate for pairs of photons, cf. 
section \ref{A}.):
\begin{equation}
\label{CH74_E3}
p(\alpha_1, \beta_1)  + p(\alpha_1, \beta_2)  
+p(\alpha_2, \beta_1) - p(\alpha_2, \beta_2) \leq p^A(\alpha_1) +p^B(\beta_1),
\end{equation} 
where $p(\alpha_i, \beta_j) = n_{oo}(\alpha_i, \beta_j)/N, p^A(\alpha_1) = S_o^A(\alpha_1)/N, p^B(\beta_1)=S_o^B(\beta_1)/N.$

However, as pointed out by Clauser and Horne \cite{CH74}, this inequality suffers from the  problem that 
the number $N$ and, hence, probabilities, are not well-determined in an  experiment. To solve this problem, 
 (\ref{CH74_E3}) can be transformed into the inequality:
\begin{equation}
\label{CH74_P}
T = \frac{p(\alpha_1, \beta_1)  + p(\alpha_1, \beta_2)  
+p(\alpha_2, \beta_1) - p(\alpha_2, \beta_2)}{p^A(\alpha_1) +p^B(\beta_1)} \leq  1.
\end{equation} 
(We call this inequality \cite{CSH} {\it CH-inequality for ratio of probabilities}, see Remark 1 later.)  
And finally $T$ can be represented as ratio of detection count rates:
\begin{equation}
\label{CH74_R}
T = \frac{R(\alpha_1, \beta_1)  + R(\alpha_1, \beta_2)  
+R(\alpha_2, \beta_1) - R(\alpha_2, \beta_2)}{R^A(\alpha_1) +R^B(\beta_1)} \leq  1,
\end{equation} 
where $R(\alpha, \beta)$ and $R^A(\alpha_1), R^B(\beta_1)$ are coincidence and single rates, respectively. 
(Following Clauser and Shimony \cite{CSH} we call this inequality  {\it CH-inequality for ratio 
of detection count rates} or simply the {\it ratio-rates CH-inequality}, see again Remark 1.)
This inequality is evidently equivalent to the following inequality in Eberhard's notation, 
i.e., with the total numbers of coincidences and single counts, instead of the rates: 
\begin{equation}
\label{CH74_Eo}
T = \frac{n(\alpha_1, \beta_1)  + n(\alpha_1, \beta_2)  
+n(\alpha_2, \beta_1) - n(\alpha_2, \beta_2)}{S^A(\alpha_1) +S^B(\beta_1)} \leq  1.
\end{equation} 
(We call this inequality  {\it CH-inequality for ratio 
of detection counts} or simply the {\it ratio-counts CH-inequality}, see again Remark 1.)

This inequality can be directly derived from (\ref{CH74_E1}) without relying on making any 
statements or assumptions about $N.$
We can also proceed another way around and derive the E-inequality in the form (\ref{CH74_E1}) from the
CH-inequality in the form (\ref{CH74_Eo}). Thus {\it these two inequalities are algebraically equivalent.} The problem 
of their statistical equivalence will be studied in section \ref{stat1}. And it is more complicated.

\medskip

{\bf Remark 1.} (On terminology) The CH-inequality 
for probabilities, see (\ref{CH74_E3}), is sometimes called simply the CH-inequality.
At the same time, in the  paper of Clauser and Shimony  \cite{CSH} this inequality 
was considered as just an intermediate  step towards the ratio-rates CH-inequality, 
see (\ref{CH74_R}). Thus it would be natural to refer to 
the latter as the CH-inequality. This terminology is used, e.g., 
by Shimony \cite{SHI}, who called the ratio-rates inequality, the BCH-inequality,  
see also the experimental proposal of Fry  and Walther \cite{Fry}.
 The main source of referring to (\ref{CH74_E3}) as 
the CH-inequality is  that the material in the original CH-paper  \cite{CH74} was presented 
in a very compact form; in particular, the inequalities for ratios of probabilities and rates, see (\ref{CH74_P}) and 
(\ref{CH74_R}), were not written anywhere. The authors just remarked that the upper limit in (\ref{CH74_E3})
can be experimentally testable without $N$ being known. This statement can be interpreted as simply 
the (algebraic) equivalence of the inequality (\ref{CH74_E3}) to the E-inequality (\ref{CH74_E1}) achieved 
with multiplication of the right- and left-hand sides of (\ref{CH74_E3}) by $N$
(again under the assumption of statistically constant production rate). 
\medskip

\subsection{Application of Chebyshev's inequality to data from the Vienna-test}  
\label{YYY}

Consider statistical data that is normally distributed. 
The information about the mean value $\mu$ and the standard deviation $\sigma$ is sufficient to find the 
spread of these data relative to 
the number of standard deviations from the mean value. Denote the mean value and the standard deviation by 
the symbols $\mu$ and $\sigma,$ respectively.
It is known that 
68\% of these data are within $1\sigma$-deviation from $\mu$, 
95\%  of the data are within $2\sigma$-deviations from $\mu$, 
and approximately 99\% of the data are within $3\sigma$-deviation 
from $\mu.$

However,  if the statistical data set is not normally distributed, i.e., its density
  deviates from the bell shape, 
then a different amount could be within $k \sigma$-deviation, $k=1,2,3...$. In this case
one can apply {\it Chebyshev's inequality} \cite{Tchebichef} -- a powerful tool   
to get to know what fraction of the statistical data falls within a few standard deviations 
from the mean value. We recall that the Chebyshev theorem states that, for any random variable $\xi$ with finite second 
moment\footnote{In particular, the Chebyshev inequality is applicable to any bounded random variable.}, i.e.,  
$E (\vert \xi\vert^2) < \infty,$ where $E$ denotes the expectation value,
and any positive number  $c,$
\begin{equation}
\label{CHEB}  
P(\vert \xi\vert \geq c) \leq \frac{E (\vert \xi\vert^2)}{c^2}.
\end{equation}
Typically in applications one starts with a random variable $\eta$   
and in (\ref{CHEB}) selects $\xi= \eta -\mu,$ where   $\mu=E (\eta)$ is  the mean value of $\eta.$ Thus
\begin{equation}
\label{CHEB1}  
P(\vert \eta - \mu\vert \geq c) \leq \frac{E (\vert \eta -\mu \vert^2)}{c^2}=\Big(\frac{\sigma}{c}\Big)^2,
\end{equation}
where $\sigma$ is standard deviation of $\eta.$
We remark that a violation with $c=k \sigma$ results in confidence probability for a violation of $(1-1/k^2).$

By using the values for the mean value and standard  deviation calculated in \cite{Zeilinger} and by applying Chebyshev's inequality,  
we can find (without knowing the probability distribution exactly, but only assuming that the dispersion is finite, see footnote 9) that 
the E-inequality is violated statistically significantly. 

\medskip

{\bf Remark 2.} It is important to present the procedure 
of calculation of the  empirical values $m_J$ and $s_J,$ which was used by Giustina {\it et al.}
 \cite{Zeilinger}: ``After recording  for a total of  300 $\rm{s}$ per setting we divided our data
 into 10-$\rm{s}$ blocks and calculated  the standard deviation  of the resulting 30 different 
 $J$ values.'' 

\medskip

We consider random variables corresponding to 
the left-hand and right-hand sides of the inequality (\ref{CH74_E1}):
\begin{equation}
\label{MMM1}
J_{\rm{pair}}= n_{oo}(\alpha_1, \beta_1)  + n_{oo}(\alpha_1, \beta_2)  
+n_{oo}(\alpha_2, \beta_1) - n_{oo}(\alpha_2, \beta_2)
\end{equation}
and 
\begin{equation}
\label{MMM2}
J_{\rm{single}}=S_o^A(\alpha_1) +S_o^B(\beta_1).
\end{equation}
Thus
\begin{equation}
\label{MMM3}
J=  J_{\rm{single}} - J_{\rm{pair}} .
\end{equation}
Denote the mean value and standard deviation of the random variable $J$  by the symbols $\mu_J$ and $\sigma_J,$ respectively.\footnote{We remark 
that, since, for experimental runs of fixed duration, the number of emitted photon pairs is a bounded random variable,    
the random variable $J$ is bounded and, hence,   $E\vert J\vert^2 <\infty.$ Therefore dispersion is well defined and the Chebyshev inequality 
is applicable.}  
Thus $\mu_J= E (J)$ and $\sigma_J^2= E(J - \mu_J)^2.$ 
Our aim is to estimate the confidence interval for the mean. 
As always we shall use the statistical estimates of the mean and dispersion:
\begin{equation}
\label{MMM}
\bar{J}= \frac{1}{L} \sum_{i=1}^L J_i, \;  
s_J^2 = \frac{1}{L-1} \sum_{i=1}^L (J_i - \bar{J})^2
\end{equation}
(in the experiment \cite{Zeilinger} $L=30,$ this is the number 
of the 10-$\rm{s}$ blocks, see Remark 2, each block is used to calculate $J_i,$ the number of pairs in each block is very large and
in such a framework its exact number 
is not so important).


We shall also use {\it the standard error of the mean} (the standard deviation of the sample-mean's estimate of a population mean):
\begin{equation}
\label{MMM4}
\rm{SE}_{\bar{J}}= \frac{s_J}{\sqrt{L}}.
\end{equation}
and {\it the standard deviation of the mean}:
\begin{equation}
\label{MMM5}
\rm{SD}_{\bar{J}}= \frac{\sigma_J}{\sqrt{L}}.
\end{equation}  
We remark that $\rm{SE}_{\bar{J}}$ decreases as $\sqrt{L}$ with increase of the size of the sample $L.$
\footnote{This makes intuitively also a lot of sense: the larger the sample one has the smaller the confidence interval for the mean value.
At the same time one must not overestimate the role of getting a very small standard error of the mean. It mainly means that one was able 
to perform measurements for very long runs of the experiment and, in particular, to guarantee the stability of functioning of the source
and the measurement devices.}

By using the Chebyshev inequality  we obtain:
\begin{equation}
\label{MMM5}
P(\vert \bar{J} - \mu_J\vert \geq c)\leq \frac{\rm{SD}^2_{\bar{J}}}{c^2}.
\end{equation}
We proceed by using the standard error of the mean, instead of  the standard deviation of the sample mean
(in the formal mathematical presentation one has to use the correction related to the finite $L,$ see \cite{Saw}, \cite{Kaban}  for details):
\begin{equation}
\label{MMM5}
P(\vert \bar{J} - \mu_J\vert \geq c)\leq \frac{\rm{SE}^2_{\bar{J}}}{c^2}.
\end{equation}
From \cite{Zeilinger} we take the values $\bar{J} \approx -4224$  and 
$\rm{SE}_{\bar{J}} \approx 61.23.$ 
This is a $> 60 \sigma$ violation, where $\sigma\equiv \rm{SE}_{\bar{J}}.$ However, 
as was pointed out, for the statistical data from the Vienna test one cannot assume 
that these data are normally distributed.
Therefore further analysis is needed. By using the the Chebyshev inequality 
we estimate the confidence interval corresponding to the confidence level 99.95\%. (In applied statistics already the level 95\% is considered 
as sufficiently high.)
Here $c\approx 2738.$ Thus:
\begin{equation}
\label{AAAB}  
P(m_J \in [\bar{J} - 2738, \bar{J} +2738]) = P(m_J \in [-6962, -1486]) \geq 0.9995.
\end{equation}
Thus the confidence that can be placed in the result of the Vienna-test is very high.
The demonstrated violation of the E-inequality cannot be a matter of chance.

\subsection{Statistical (non-)equivalence}
\label{stat1}

Generally (i.e., without additional assumptions on probability distributions) the algebraic equivalence 
of two tests does not imply their statistical equivalence, at least for finite samples.
In particular, the tests based on the E-inequality and the CH-inequality 
in the form (\ref{CH74_Eo})
are not statistically equivalent. The latter means that violation of one of them with $k \sigma,$ where $k$ is sufficiently 
large, need not imply that another will be violated 
with the same $k.$ It may happen that the significance of the violation changes essentially.  This is a general statistical feature, 
i.e., it is not coupled rigidly with the two statistical tests under consideration, see appendix.

We remark that if the conditions of the central limit theorem are satisfied (in particular, 
for identically distributed independent random variables), then by using the $\delta$-method \cite{Cramer} 
(error propagation method) we one can prove the statistical equivalence of the E-test and the CH-test 
for $L\to \infty.$ However, for finite $L,$ in general it is true that by looking at different functions of 
statistics of interest and using  the 
delta method, one   
can get any answer one likes. All of these answers are just approximations, and 
some approximations are better than others. As was pointed out, for  $L\to \infty,$ 
they give the same answer,  but for fixed $L$ they all give different answers.

\subsection{Statistically significant violation of the CH-inequality (in the ratio form) for the Vienna-test}

Thus on the basis of purely theoretical arguments one cannot derive   a
statistically  significant violation of the CH-inequality in the form (\ref{CH74_Eo})
from statistically  significant violation of the E-inequality. One has to use 
again the experimental data.

In this section we apply Chebyshev's inequality to show
that the statistical data collected in the Vienna-test also implies statistically significant 
violation of the (ratio-counts) CH-inequality.  Set $\mu_T=E (T)$ (the mean value of the random variable $T)$
and $\sigma_T^2= E(T- \mu_T)^2$ (its dispersion). 
We shall use the statistical estimates of the mean and dispersion:
$\bar{T}= \frac{1}{L} \sum_{i=1}^L T_i, \;  
s_T^2 = \frac{1}{L-1} \sum_{i=1}^L (T_i - \bar{T})^2$
(in the experiment \cite{Zeilinger} $L=30,$ this is the number 
of the 10-$\rm{s}$ blocks, each block is used to calculate $T_i)$. 
We also consider the standard deviation of the mean $\rm{SD}_{\bar{T}}= \frac{\sigma_T}{\sqrt{L}}$
and and the standard error of the mean $\rm{SE}_{\bar{T}}= \frac{s_T}{\sqrt{L}}.$

By using  the  data from \cite{Zeilinger} and more detailed presentation in \cite{Zeilinger1} we obtain
$\bar{T} \approx  1.0394$ and $\rm{SE}_{\bar{T}} \approx 0.0006.$ 
This yields  a $> 60 \sigma$ violation, where $\sigma= \rm{SE}_{\bar{T}}.$
From the Chebyshev inequality 
\begin{equation}
\label{MMM5A}
P(\vert \bar{T} - \mu_T\vert \geq c)\leq \frac{\rm{SE}^2_{\bar{T}}}{c^2}.
\end{equation}
we estimate the confidence interval corresponding
to the confidence level 99.95\%. We have $c\approx 0.027.$
\begin{equation}
\label{AAAB}  
P(m_T \in [\bar{T} - 0.027, \bar{T} +0.027]) = P(m_T \in [1.0124, 1.0664]) \geq 0.9995.
\end{equation}
Thus the confidence that can be placed in the result of the Vienna test is very high.
The demonstrated violation of the (ratio-counts) CH-inequality cannot be a matter of chance.

\section{The Vienna-test for the  CH-inequality (in the ratio form): taking into account intensity drift}
\label{A}

In all previous considerations we assumed, as Eberhard originally \cite{Eberhard}, that the number 
of emitted pairs $N$ is constant during the experiment and does not depend 
on angles $(\alpha_i, \beta_j).$ In the real experiment, the intensity drift was very small  \cite{Zeilinger}.
In \cite{Zeilinger1}, a data analysis procedure was proposed being based on the following
assumption:
$$
N(\alpha_1, \beta_1)/N(\alpha_2, \beta_1)=
S_o^B(\alpha_1, \beta_1)/S_o^B(\alpha_2, \beta_1),....
$$
Then one proceeds not simply with coincidence and single counts 
$n_{oo},  n_{oe},...,\\ S_o^A, S_o^B,$  but with their normalized values based on  
the above mentioned proportion of intensities. Denote normalized quantities as $\tilde{n}_{oo},
\tilde{n}_{oe},...,
\tilde{S}_o^A, \tilde{S}_o^B.$ For such normalized numbers of
coincidences and singles, we can use the E-inequality:
\begin{equation}
\label{CH74_Ev}
\tilde{n}_{oo}(\alpha_1, \beta_1)  + \tilde{n}_{oo}(\alpha_1, \beta_2)  
+\tilde{n}_{oo}(\alpha_2, \beta_1) - \tilde{n}_{oo}(\alpha_2, \beta_2) 
\leq \tilde{S}_o^A(\alpha_1) +\tilde{S}_o^B(\beta_1)  .
\end{equation} 
And we jump directly to the CH-inequality for the total numbers of coincidences and singles 
(which is equivalent to the CH-inequality for the rates):
\begin{equation}
\label{CH74_Eov}
T^\prime = \frac{\tilde{n}_{oo}(\alpha_1, \beta_1)  + \tilde{n}_{oo}(\alpha_1, \beta_2)  
+\tilde{n}_{oo}(\alpha_2, \beta_1) - \tilde{n}_{oo}(\alpha_2, \beta_2)}{\tilde{S}_o^A(\alpha_1) 
+\tilde{S}_o^B(\beta_1)} \leq  1.
\end{equation} 
Using data collected in \cite{Zeilinger} and  \cite{Zeilinger1},  
we get   $\bar{T^\prime} \approx 1.0384$  with a violation  $>60 \sigma$ which shows that also 
the experimental data taking into account intensity drift lead to the same amount of non-classical 
correlations.  

\section{Concluding remark}

The statistical data collected in the Vienna-test \cite{Zeilinger, Zeilinger1}  
violated statistically significantly not only the 
E-inequality, but equivalently also the CH-inequality for ratio of detection counts and, hence,
the ratio-rate inequality, cf. \cite{Kwiat1}, \cite{Kwiat2}. Thus, we can consider 
this experimental test as closing the detection loophole also with regard to 
the CH-inequality in the ratio-form. 

\section{Appendix: Statistical non-equivalence of (algebraically equivalent) linear and ratio test}

Let $X$ be a set of data sampled from realizations of some random variable $x.$
Take two (for a moment arbitrary positive valued) functions, $J_1(x)$ and $J_2(x);$  
set $J(x) =J_1(x) - J_2(x)$ and $T(x)= \frac{J_2(x)}{J_1(x)}.$ Consider two tests for 
statistical data:
\begin{equation}
\label{Q1}
 J(x) \geq 0;
\end{equation}
\begin{equation}
\label{Q2}
 T(x) \leq 1.
\end{equation}
Suppose that the data $X$ showed $k \sigma_J$ violation of the first inequality, where $k$ is large. Thus violation 
of this inequality is significant. 
Our aim is to show that the same data can in principle show insignificant violation of the second inequality, say $\gamma \sigma_T,$
where $\gamma$ is very small.

Suppose, for example, that the data $X$ was obtained as the result of measurements of 
 a discrete random variable $x=x_1, x_2,$ where $x_1, x_2$ are two arbitrary real numbers. It takes these values with probabilities
$p_1$ and $p_2=1-p_1.$ Here we need to set the values of all functions only in the two points, $x_1$ and $x_2:$
$$ 
J_1(x_i)=A_i, J_2(x_i)=B_i,  A_i,B_i >0, i=1,2.
$$
We take $B_i = (1+\epsilon_i) A_i, \epsilon_i >0.$ We have: 
$$
\mu_J=-(\epsilon_1 A_1 p_1 + \epsilon_2 A_2 p_2), \sigma_J^2= p_1 p_2 (\epsilon_1 A_1- \epsilon_2 A_2)^2;
$$
$$
\mu_T= 1+ \epsilon_1 p_1 + \epsilon_2 p_2,  \sigma_J^2= p_1 p_2 (\epsilon_1 - \epsilon_2 )^2;
$$
$$
R_J= \frac{- \mu_J}{\sigma_J}= \frac{\epsilon_1 A_1 p_1 + \epsilon_2 A_2 p_2}{ \sqrt{p_1 p_2} \vert \epsilon_1 A_1- \epsilon_2 A_2\vert};
\; \; R_T= \frac{\mu_T-1}{\sigma_T}= \frac{\epsilon_1  p_1 + \epsilon_2 p_2}{ \sqrt{p_1 p_2} \vert \epsilon_1 - \epsilon_2 \vert}.
$$
By playing with parameters we want to make $R_J >>1$ and at the same time $R_T<<1.$ First we set $\epsilon_1= \lambda \epsilon_2,$
where $\lambda >1.$      
We have 
$$
R_T = \sqrt{\frac{p_1}{p_2}} + \frac{1}{\sqrt{p_1 p_2} (\lambda- 1)}.
$$  
To make the first term very small,
we select $p_1= \delta^2 <<1,$ so $p_2\approx 1;$ to make the second term very small, we select $\lambda$ in such a way 
that its denominator is very large, i.e.,  $\sqrt{p_1 p_2} (\lambda- 1) >> 1.$ Thus, for the model parameters satisfying 
conditions 
\begin{equation}
\label{Q3}
\delta <<1, \; \lambda >> 1/\delta,
\end{equation}
the inequality (\ref{Q2}) is violated insignificantly. We remark that, since the parameter $\epsilon_2$ has not yet been 
constrained, it is possible to make the absolute value of expectation very large, $\mu_T>>1.$  
We now want to make $R_J>>1.$ We represent $A_1= aA_2,$  and we have:
$
R_J== \frac{a \lambda p_1 +  p_2}{ \sqrt{p_1 p_2} \vert a \lambda- 1\vert}.
$
Now we select $a$ as
\begin{equation}
\label{Q4}
 a \lambda - 1> 0.
\end{equation}
Thus 
$$
R_J== \frac{a \lambda p_1 +  p_2}{ \sqrt{p_1 p_2} (a \lambda- 1)} = 
\sqrt{\frac{p_1}{p_2}} + \frac{1}{\sqrt{p_1 p_2} (a\lambda- 1)}.
$$
The first summand is negligibly small, see (\ref{Q3}), it does not play any role in our considerations. The parameter $a$ has 
to be selected in such a way that the second summand will be very large. Thus $\sqrt{p_1 p_2} (a\lambda- 1)<<1,$ or by taking into account 
(\ref{Q4})  we obtain that $0<a\lambda - 1<< 1/ \sqrt{p_1 p_2} \approx 1/ \delta.$  Take very large natural number $k.$ Suppose 
that 
\begin{equation}
\label{Q5}
a  = 1/\lambda+ 1/k \delta \lambda.
\end{equation}
Then $R_J\approx k.$\footnote{For example, set $\delta=0.1,$ i.e., $p_1=0.01, p_2=0.99.$ We remark that to get very small $R_T$ and at 
the same time very large $R_J$ the probability distribution of our model has to be strongly asymmetric. Then $\lambda=102$ guarantees 
that $R_T <0.2.$ Finally, by setting $k=69$ we obtain that it is sufficient to take $a=0.0112.$}   
    
\bigskip 

{\bf Acknowledgment:} 
We would like to thank all authors of the original experiment-paper \cite{Zeilinger}  as they contributed to the data we used in this paper;
especially, to M. Giustina whose numerous comments and suggestions essentially improved the paper. 
We would like to thank R. Gill,  
and R. Pettersson for discussions on statistical aspects of the paper and J.-A.
Larsson and G. Adenier for discussions on the role of the fair sampling assumption in 
derivation of various versions of Bell inequality. 
This work was partially supported by MPNS COST Action MP1006, Fundamental Problems in Quantum Physics (I. Basieva), a visiting fellowship 
(A. Khrennikov) to  
Institute for Quantum Optics and Quantum Information, Austrian Academy of Sciences,
 and   EU Marie-Curie Fellowship (S. Ramelow), PIOF-GA-2012-32985.

\end{document}